\documentclass[twocolumn,preprintnumbers,amsmath,amssymb,floatfix]{revtex4}
\usepackage{graphicx}
\usepackage{dcolumn}
\usepackage{bm}
\graphicspath{{./},{./figs/}}
\usepackage[latin1]{inputenc}
\begin{document}


\title{Charging energy spectrum of black phosphorus quantum dots}

\author{M. A. Lino$^{1,2}$, J. S. de Sousa$^2$, D. R. da Costa$^2$, A. Chaves$^2$, J. M. Pereira$^2$, G. A. Farias$^2$}
\affiliation{$^{1}$Universidade Federal do Piauí, Departamento de Física, CEP 64049-550, Teresina, Piauí, Brazil}
\affiliation{$^2$Universidade Federal do Ceará, Departamento de Física, Caixa Postal 6030, 60455-760 Fortaleza, Ceará, Brazil}

\begin{abstract}
We present a theoretical study of the charging effects in single and double layer black phosphorus quantum dots (BPQDs) with lateral sizes of 2 nm and 3 nm. We demonstrate that the charging of BPQDs are able to store up to an $N_{max}$ electron (that depends on the lateral size and number of layers in the QD), after which structural instabilities arises. For example, 3 nm wide hydrogen-passivated single layer BPQDs can hold a maximum of 16 electrons, and an additional electron causes the expelling of hydrogen atoms from the QD borders. We also calculated the additional energy ($E_A$) spectrum. For single layer QDs with 2 and 3 nm of lateral sizes, the average $E_A$ is around 0.4 eV and 0.3 eV, respectively. For double layer QDs with the same sizes, the average $E_A$ is around 0.25 eV and 0.2 eV, respectively.
\end{abstract}

 \maketitle



The research on graphene motivated the study of many classes of  bidimensional materials like silicene, germanene, phosphorene, metallic oxides, boron nitride, and dicalcogenates of transition metals (MoS$_2$, TiS$_2$, WS$_2$) \cite{ref1,ref2, ref3}. In particular, black phosphorus (BP) is a layered allotrope of phosphorus that can be exfoliated into single or few layers. The interesting physical and chemical properties of BP compared to other 2D materials are due to its unique direct band gap of the order of 2~eV for the monolayer and 1.88 eV for the bilayer\cite{ref4,ref5,ref6,ref7,ref8}, high carrier mobilities, and highly anisotropic band structure \cite{xia2014,castellanos2015,rudenko2014,cakir2014,pereira2015}. A recent study has shown that the large binding energy of excitons in phosphorene can withstand in-plane eletric fields as high as 200 kV/cm, giving rise to excited excitonic states \cite{chaves2015}.  Due to the  band structure anisotropy, the excitons are elongated in the armchair direction, where the effective masses are lower  \cite{chaves2016}. Exquisite BP based nanostructure have also been theoretically predicted. Density Functional Theory (DFT) calculations have shown that phosporene nanoribbons with armchair (zigzag) borders exhibit indirect (direct) bandgap semiconductor \cite{ref9}, while single walled phosphorene nanotubes are direct bandgap semiconductures for both armchair and zigzag chiralities  \cite{ref10}. Those predictions have yet to be confirmed experimentally.

Recently, some groups have successfully obtained BP quantum dots (BPQDs). Sofer \textit{et al}  produced BPQDs with few layers with average of 15 nm and large bandgaps  \cite{ref12}. The wet exfoliation method of Sun \textit{et al} produced BPQDs as small as $2.6\pm1.8$ nm of diameter, and  $1.5\pm0.6$ nm of thickness \cite{ref13}. Xu \textit{ et al.} also produced BPQDs with average size of $2.1\pm0.9$ nm in large scale by solvothermal synthesis \cite{ref14}. Zhang et al. produced BPQDs by means of wet exfoliation, obtaining BPQDs with lateral sizes of $4.9\pm1.6$ nm and thicknesses of $1.9\pm0.9$ nm \cite{ref11}. As a proof-of-concept of the confinement properties of their BPQDs, they were mixed in a polyvinylpyrrolidone (PVP) polymer matrix to produce the active layer of  low power non-volatile rewritable memory devices. They reported a high ON/OFF current ratio of the order of $6.0 \times 10^4$ at reading voltages of only 0.2 V, which is significantly higher than C$_{60}$ and MoS$_2$ based PVP devices.

Although the exact write/erase mechanism of the nonvolatile memories produced by Zhang \textit{et al.} were not elucidated in their work, their device is analogous to the well known semiconductor nanocrystals (NC) nonvolatile flash memories \cite{desousa2011,peibst2010,desousa2002}. The operation of NC memory devices consists in storing (programming), holding (retention), and removing (erase) charges in the NCs. The low dimensionality of the NCs imposes behavior almost entirely governed by quantum mechanics: (i) quantum tunneling is the basic mechanism for write/erase operations, and (ii) quantum confinement determines the conditions for charge retention in the NCs. Thus, it is reasonable to assume that the BPQD memory device of Zhang \textit{et al.} makes use of the same mechanisms.  As shown by de Sousa et al., in these systems many electrons are typically stored in a single write pulse \cite{desousa2011}. Since electrons are confined in a very small volume, as more electrons are confined, more energy becomes necessary to add an extra electron.   Disregarding exchange-correlation effects, the total energy in a semiconductor QD containing $N$ electrons is given by the sum of its single-particle energies $\varepsilon_i$ plus the electrostatic energy $U(N)$, such that $E_{tot}(N) = \sum_{i=1}^N \varepsilon_i + U(N)$. The electrochemical potential of the QD containing $N$ electrons is given by $\mu(N) = E_{tot}(N)-E_{tot}(N-1)$, and the energy necessary to add an additional electron to the QD is defined as $E_A(N) = \mu(N+1) - \mu(N)$  \cite{thean2001}.

\begin{figure}
\includegraphics[scale=0.45,clip=true]{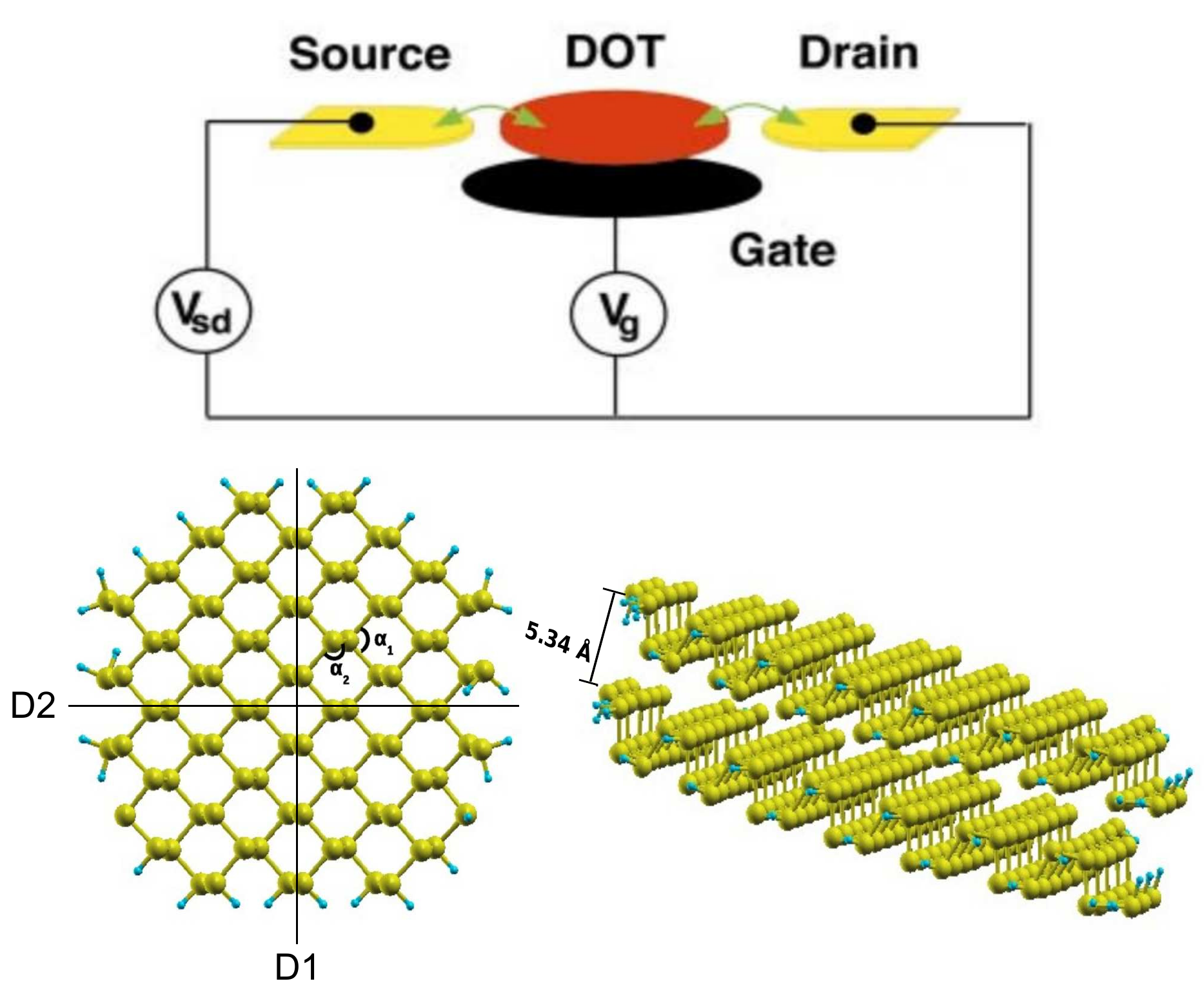}
\caption{\label{set} (top) Schematics of QD based SET. The QD is couple to the source and drain reservoirs  though tunneling barriers \cite{ref-review}. The gate contact controls the local energy reference in the QD. (bottom) Top  view of single layer and lateral view of double layer PBQDs investigated in this work. The dangling bonds in the QD borders are saturated with H. Diameters D1 and D2 are measured in the zigzag and armchair directions, respectively.}
\end{figure}

The single (few) electron transistor (SET) is a model device whose working principle relies entirely on the Coulomb blockade effect  \cite{ref15}. A SET is essentially a quantum island weakly coupled to reservoirs by tunnelling barriers, as shown in Figure \ref{set}. The conductance across the quantum island depends on the relative alignment of the chemical potential of the source, drain and the quantum island. The total energy of the quantum island is controlled by the number of confined electrons   and by the gate voltage $V_g$. Keeping the drain-to-source voltage $V_{ds}$ constant, one can control $N$ (and $\mu(N)$ as well) by raising/lowering $V_G$. For a suitable alignment of $\mu(N)$ as compared to $\mu_{s,d}$, the conductance $dI/dV_g$ as a function of $V_g$ will exhibit peaks, for $V_g$ in-between peaks, the conductance across the QD is turned off. SET theory summaries three fundamental conditions to observe Coulomb blockade effect in the electrical transport through small quantum islands:  (i) the single electron charging energy must be much larger than the thermal energy ($E_A >> 2 k_BT$), (ii) the drain-to-source voltage should be of the order of the charging energy and must obey $e V_{ds} < 2 E_A$, (iii) the tunnel coupling between the quantum islands and the leads has to be small to ensure a long lifetime $\Delta t$ of the electrons in the quantum island  such that the uncertainty in energy $\Delta E \approx h /\Delta t$ must not exceed $E_A$. 

In this work, we calculate the feasibility of producing SETs from BPQDs by calculating the additional energy spectrum of single and double layer isolated BPQDs as small as 2 nm and 3 nm by means of \textit{ab initio} calculations.  We  demonstrate that single and double layer BQPDs with diameters up to 3 nm exhibit additional energy of the order of 0.2 eV, much above the thermal fluctuation energies, and should therefore exhibit Coulomb blockade effect even at room temperature. 

Our first-principles methodology is based on the Density Functional Theory (DFT) \cite{kohn} as implemented in the Siesta program \cite{soler96,soler2002}. We make use of the Generalized Gradient Approximation (GGA) for the exchange-correlation functional \cite{perdew96} and norm-conserving Troullier-Martins pseudopotentials \cite{troullier91} in Kleinman-Bylander factorized form \cite{Kleinman82}. As for the basis set, we use double-zeta basis function composed of numerical atomic orbitals of finite range augmented by polarization functions (DZP basis set). The fineness of the real-space grid integration was defined by a minimal energy cutoff of 180 Ry. The range of each orbital is determined by an orbital energy confinement of 0.01 Ry. The geometries were considered optimized when the remanent force components were less than 0.04 eV/\AA.

To construct nearly circular BPQDs, we started with relaxed single and double layer BP bulk structures, whose  angles and atomic distances were similar to the ones reported by Morita \cite{morita-1986}. The atoms lying outside a circle with a given diameter (2 nm and 3 nm) were removed and the remaining dangling bonds were saturated with Hydrogens. This method results in QDs with borders of mixed zigzag and armchair symmetries. The BPQDs were also relaxed following the same relaxation criteria of the bulk layers. We then added electrons one by one, relaxing the structure of the charged QDs after each addition, and computed the total energy $E_{tot}(N)$ of each charged QD state, were $N$ is the number of electrons in the QD. The total energy as function of the $N$ is then used to compute the additional energy spectrum of the BPQDs.

\begin{figure}
\includegraphics[scale=0.35,clip=true]{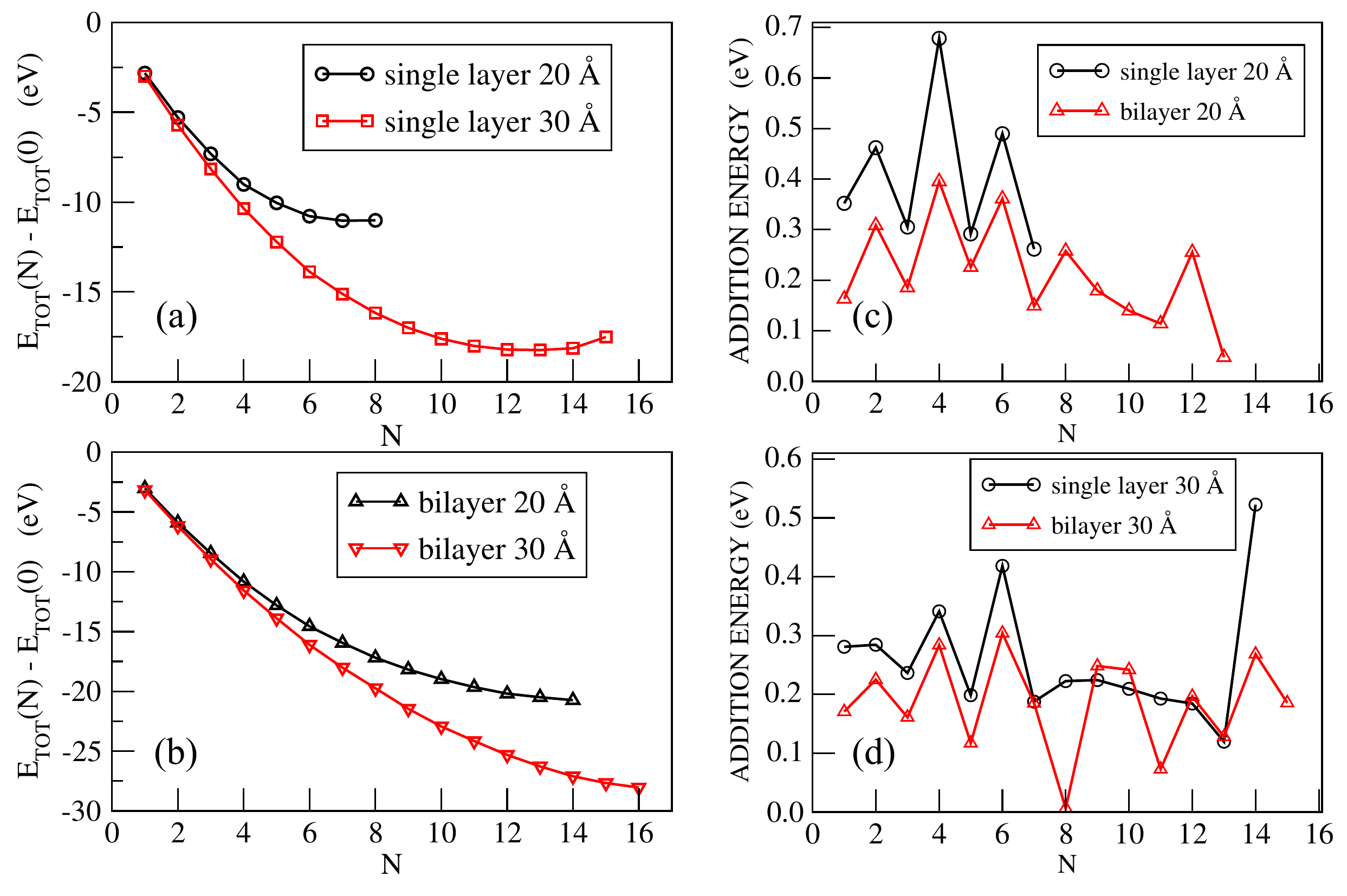}
\caption{(a-b) Total energy of single and double layer BPQDs and their additional energy spectra (c-d) as a function of the number of confined electrons.}
\label{fig:etot}
\end{figure}

\begin{figure}[ht]
\includegraphics[scale=0.5,clip=true]{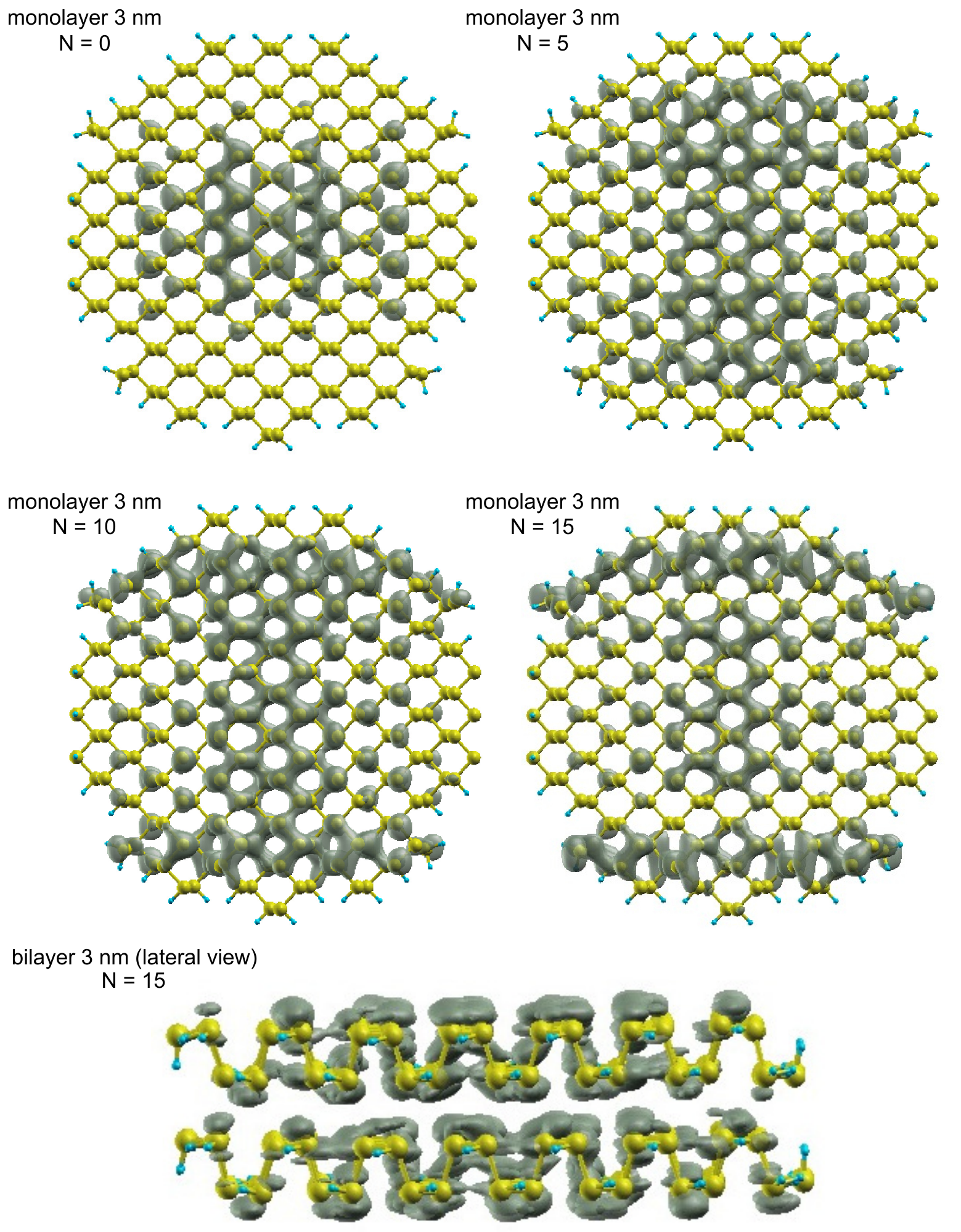}
\caption{\label{fig:density} Top view of the charge density for varying values of $N$. The lateral view of the charge density in a double layer BPQD is also shown.}
\end{figure}

 The total energy shifts $E_{tot}(N)-E_{tot}(0)$ for single and double layer BPQDs are shown in Figures \ref{fig:etot}(a-b). Interestingly, the total energy decreases as $N$ increases up to an $N_{max}$, after that the total energy increases if one continues adding electrons. The value of $N_{max}$ seems to be dependent on the lateral QD size and number of layers. For single layer QDs, $N_{max} = 7,13$ for diameters of 2 and 3 nm, respectively. For double layer QDs, Figure \ref{fig:etot}(b) suggests that $N_{max} = 14,16$, for  diameters of 2 and 3 nm, respectively. Figure \ref{fig:density} shows the charge density as a function of N for single layer BPQDs. As $N$ increases, charges accumulate at some of the edges of the QDs. As $N$ increases, one can clearly observe the effect of the band structure anisotropy on the charge density. We remark that we added to each structure more electrons than what is shown in Figures \ref{fig:etot}(a-b) to observe the behavior of the $E_{tot}(N)$ curve for $N>N_{max}$. However, the relaxation convergence becomes highly unstable for $N>N_{max}$, and we disregarded those data points. Such instabilities occur because the confined electrons seek to minimize the electrostatic repulsion by moving to the far edges of the QDs (which are not perfectly circular), and the accumulated charges tend to break the QD structure by breaking either P-P or P-H bonds near the QD borders. 

The additional energy spectrum is shown in Figures \ref{fig:etot}(c-d). As previously reported in other QDs, the additional energy spectra exhibit strong fluctuations (with peaks above 0.5 eV for single layer QDs) as $N$ increases \cite{thean2001}. The additional energy is smaller for larger diameters and larger number of layers. For single layer QDs with $D = 2, 3$ nm, the average $E_A$ is around 0.4 eV (up to $N=7$) and 0.3 eV (up to $N = 14$), respectively. For double layer QDs with $D = 2, 3$ nm, the average $E_A$ is around 0.25 eV (up to $N = 13$) and 0.2 eV (up to $N = 15$), respectively.

\begin{figure}[ht]
\includegraphics[scale=0.5,clip=true]{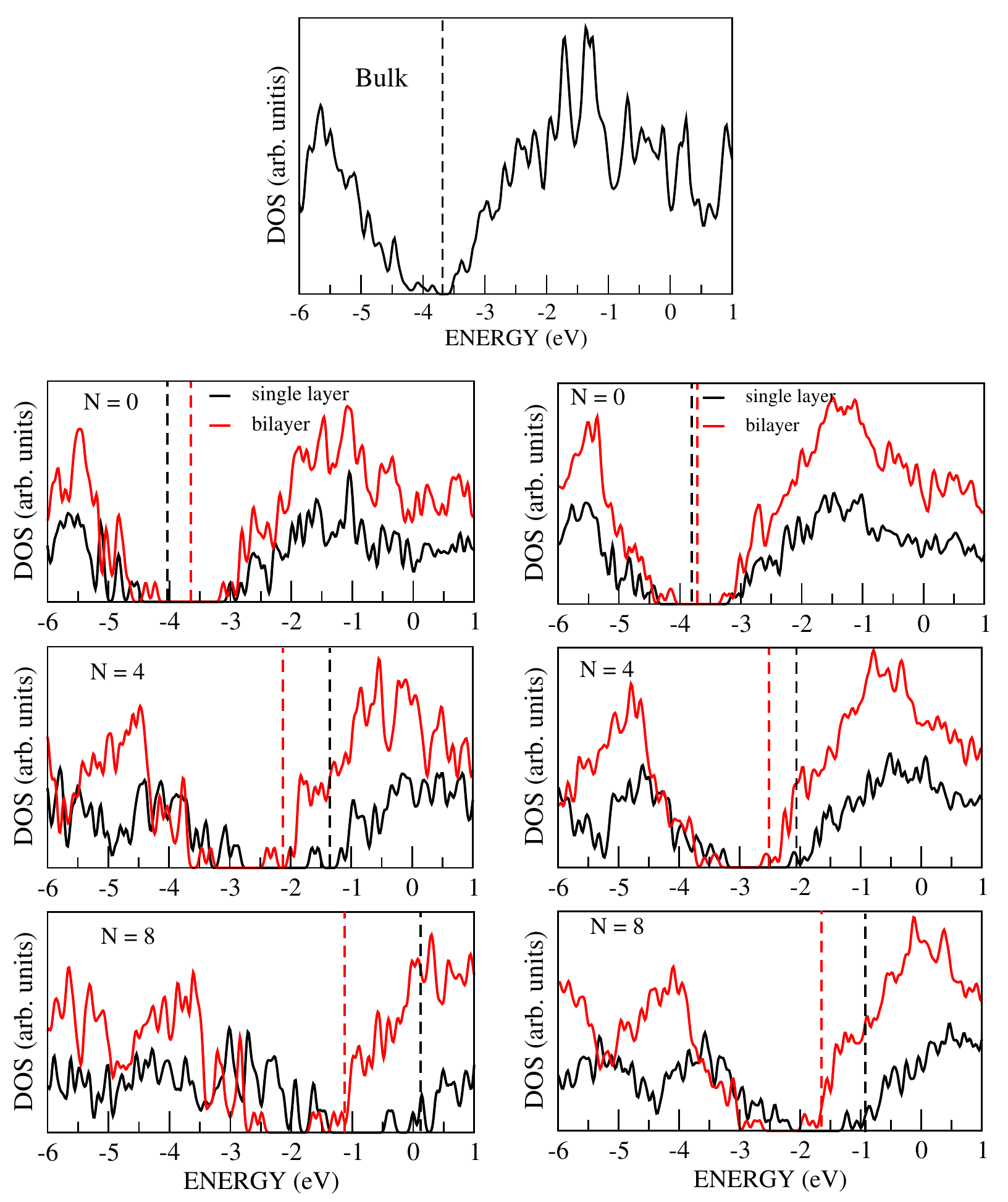}
\caption{\label{fig:dos} Total density of states (DOS) of single and double layer BPQDs for different charging states. Left and right panels show the DOS for QD with lateral sizes of 2 nm and 3 nm, respectively. For the sake of comparison, the DOS of single layer bulk BP is also shown.}
\end{figure}

The effect of multiple confined electrons in the total density of states of the BPQDs is shown in Figure \ref{fig:dos}. The addition of electrons induces a nearly rigid shift in the DOS and Fermi energies, with the shifts depending on the charging state of the QD. The Fermi energy of the single layer QD with 2 nm shifts from around -4.0 eV to -1.5 eV, and then to around 0.2 eV, for N = 0, 4 and 8, respectively. The charging state also increases the difference between the Fermi energies of single and double layers.

\begin{figure*}[ht]
\includegraphics[scale=0.6,clip=true]{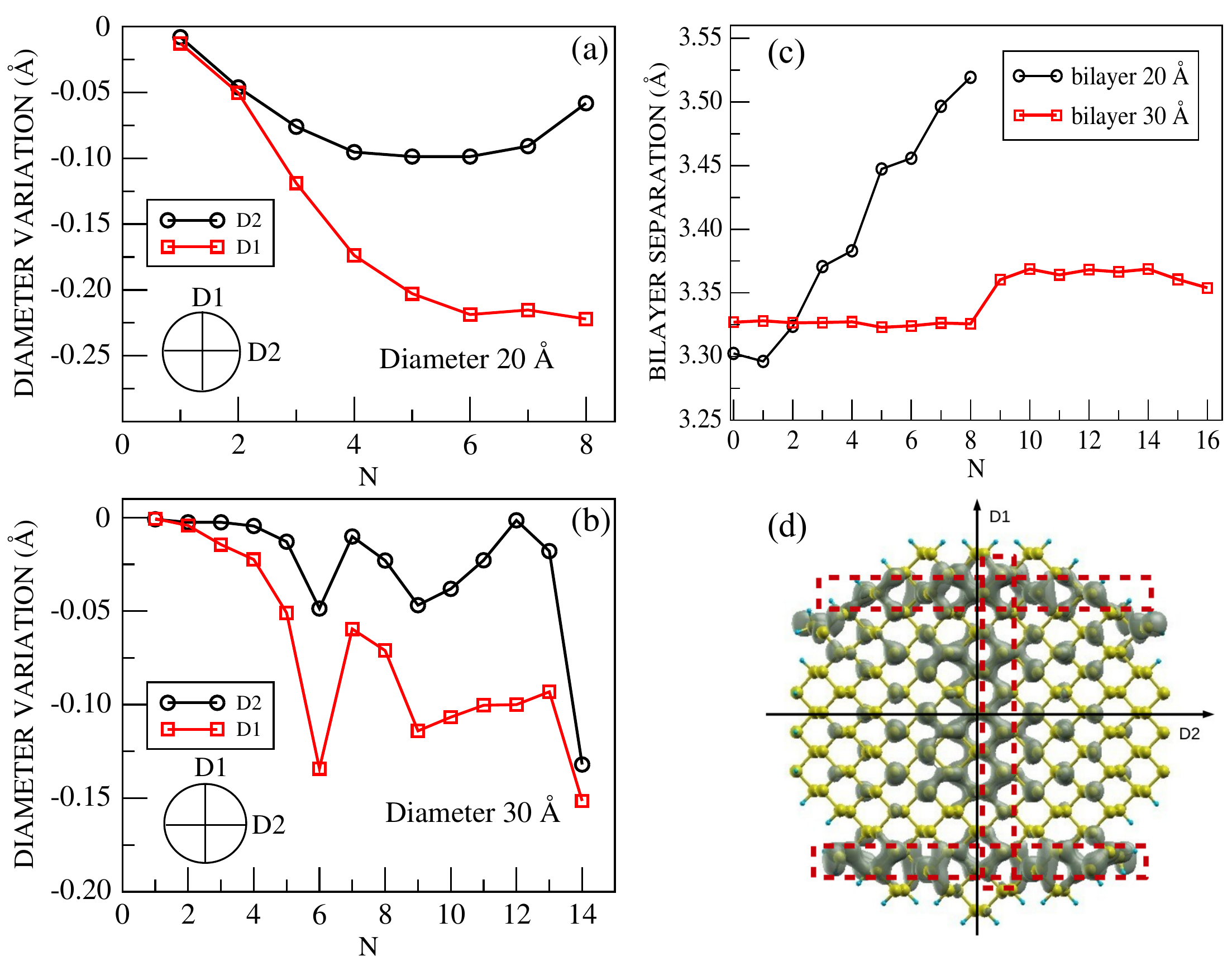}
\caption{\label{fig:structure} (a-b) Diameter variation of single layer BPQDs and (c) bilayer separation as function of the number of confined electrons. Both variations are calculated as the difference between the charged and non-charged structures. (d) Charge density for $N = 15$ in a 3 nm wide QD. The marked regions represent the regions where the swelling of the structure exhibits the highest monolayer thickness. In those regions the monolayer thickness vary between $2.3\times 10^{-3}$ nm and $3.3\times 10^{-2}$ nm. Diameters D1 and D2 are measured in the zigzag and armchair directions, respectively.}
\end{figure*}

Calculations have shown that the addition of electrons lowers the total energy of the QDs up to $N_{max}$, and subsequent charging will lead to structural instabilities. This leads to natural question of what kind of structural modification is causing the lowering of the total energy compared to the neutral QD. Since our QDs are not perfectly circular, we calculated two lateral sizes D1 (in zigzag direction)  and D2 (in armchair direction), and the bilayer separation as a function of $N$, as shown in Figure \ref{fig:structure}. For the single layer QDs, both diameters D1 and D2 are found to decrease as N increases. For up to $N=2$, D1 and D2 decrease equally. For larger values of N, we observed an anisotropic deformation of the BPQDs, with D1 decreasing twice compared to D2. The additional charges cause a swelling of the structure with a consequent increase in the thickness of the monolayer accompanied by a reduction of the diameters D1 and D2. The increase in the monolayer thickness is not uniform, it is higher in regions with increased charge densities. The charging of the QDs also induces an increase of the bilayer separation (calculated in the center of the QD). The high density of extra electrons deforms the structure as it seeks an equilibrium distribution. Due to the \emph{wrinkled} structure of BP, the more energetically favorable way to accomplish this task is either increasing the monolayer thickness or \emph{unwrinkling} the structure. Both ways increase the average separation between atoms, thereby reducing the electrostatic component of the total energy of the system. However, our calculations show that increasing the monolayer thickness is easier than \emph{unwrinkling} the structure. We conclude that the shorter in-plane bonds are stronger than the larger off-plane bonds. This gives rise to an anisotropic in-plane deformation of the structure, resulting in a larger reduction of D1 compared to D2. Besides that, the extra electrons form a mobile charge density in the conduction band. Although constrained by band structure anisotropy, electrons can move freely to occupy the QD volume, and the charge density accumulates near the edges of the QD (see Figure \ref{fig:density} for $N>0$). In those points of high charge densities, the electrostatic repulsion is strong enough to break the structure. 

Finally, our calculations also show that the average additional energy spectra of small BPQDs with one and two layers range between 0.2 eV and 0.4 eV, which is much larger than the room temperature thermal energy $k_BT$, satisfying the condition $E_A >> k_BT$ of SET theory for the observation of Coulomb blockade. We remark that our calculation was performed for isolated QDs, and the effect of a substrate may lower the additional energy, as in the case of graphene \cite{wang2012}. This topic is reserved for future study. The other two conditions, namely $e V_{ds} \approx E_A$ and relatively long lifetime of the charges in the QD, can be tailored in the design of the device itself. Therefore, we can conclude that BPQDs are excellent candidate structures to develop SET devices. Another advantage is the reasonably easy fabrication methods based on wet exfoliation of BP as reported by Zhang, Sun and Sofer \cite{ref13,ref14,ref15}. The simplicity of the fabrication of the memory device of Zhang \cite{ref13} can be easily adapted to fabricate SETs based on BPQDs.

\noindent\textbf{Acknowledgements.} The authors acknowledge the financial support from the Brazilian National Research Council (CNPq).



\end{document}